\documentclass[a4paper,11pt]{article}

\usepackage{jcappub} 

\usepackage[T1]{fontenc} 

\title{\boldmath Constraining the range of Yukawa gravity interaction from S2
star orbits}

\author[a,1]{D. Borka,\note{Corresponding author.}}
\author[b]{P. Jovanovi\'{c},}
\author[a]{V. Borka Jovanovi\'{c}}
\author[c,d,e]{and A. F. Zakharov}

\affiliation[a]{Atomic Physics Laboratory (040), Vin\v{c}a Institute
of Nuclear Sciences,\\
University of Belgrade, P.O. Box 522, 11001 Belgrade, Serbia}
\affiliation[b]{Astronomical Observatory, Volgina 7, 11060
Belgrade, Serbia}
\affiliation[c]{Institute of Theoretical and Experimental Physics,
B. Cheremushkinskaya 25, 117259 Moscow, Russia}
\affiliation[d]{Bogoliubov Laboratory for Theoretical Physics,
JINR, 141980 Dubna, Russia}
\affiliation[e]{North Carolina Central University, 1801 Fayetteville
 Street, Durham, NC 27707, USA}

\emailAdd{dusborka@vinca.rs}
\emailAdd{pjovanovic@aob.rs}
\emailAdd{vborka@vinca.rs}
\emailAdd{zakharov@itep.ru}

\abstract{We consider possible signatures for Yukawa gravity within the
Galactic Central Parsec, based on our analysis of the S2 star orbital
precession around the massive compact dark object at the Galactic
Centre, and on the comparisons between the simulated orbits in Yukawa gravity
and two independent sets of observations. Our simulations resulted in strong
constraints on the range of Yukawa interaction $\Lambda$ and showed that its
most probable value in the case of S2 star is around 5000 - 7000 AU. At
the same time, we were not able to obtain reliable constrains on
the universal constant $\delta$ of Yukawa gravity, because the
current observations of S2 star indicated that it may be highly
correlated with parameter $\Lambda$ in the range $(0 <\delta < 1)$.
For $\delta > 2$ they are not correlated. However, the same universal constant which
was successfully applied to clusters of galaxies and rotation curves
of spiral galaxies ($\delta=1/3$) also gives a satisfactory agreement with
the observed orbital precession of the S2 star, and in that case
the most probable value for the scale parameter is $\Lambda \approx 3000
\pm 1500$ AU. Also, the Yukawa gravity potential induces precession of S2 star
orbit in the same direction as General Relativity for $\delta > 0$ and for
$\delta < -1$, and in the opposite direction for $-1 <\delta < 0$. The future
observations with advanced facilities, such as GRAVITY or/and European Extremely
Large Telescope, are needed in order to verify these claims.}

\begin{document}

\maketitle

\flushbottom

\section{Introduction}
\label{sec:Sec1}

The modified theories of gravity have been proposed like alternative
approaches to Newtonian gravity on the ground of astrophysical and
cosmological consequences of the observations of the Solar system,
binary pulsars, spiral galaxies, clusters of galaxies and the
large-scale structure of the Universe. The search for non-Newtonian
gravity is part of the quest for non-Einsteinian physics
which consists of searches for deviations from Special and General
Relativity \cite{fisc99}. Different alternative gravity theories have been
proposed (see e.g. \cite{clif06,clif12,capo11a,capo11b} for reviews), such as:
MOND \cite{milg83,milg03,beke04,beke06},
scalar-tensor \cite{bran61}, conformal \cite{behn02,barb06}, Yukawa
like corrected gravity theories \cite{fisc86,fisc92,hoyl01,bere10,iori07}
and extended theories of gravity
\cite{capo02,capo03,carr04,leon11,soti10,capo11a,capo11b}.
One type of the extended theories of gravity is characterized by
power-law Lagrangians \cite{capo06,capo07}. An extension of
Post-Newtonian relativistic theory is presented by Kopeikin and Vlasov, who
used a general class of the scalar-tensor (Brans-Dicke type) theories of
gravitation \cite{cope04}.
Alternative approaches to Newtonian gravity in the framework of the weak field
limit of fourth order gravity theory have been proposed and
constraints on these theories have been discussed
\cite{zakh06,zakh07,frig07,nuci07,bork12,capo09a,iori10}.
Theories of "massive gravity" have also attracted some attention, and
they could give an exponential decay to Newton's potential at the
large distances (see e.g. \cite{ruba08} for a
review). Babichev et al. introduced a new limit of that theory, in the
weak-field approximation, which is able to capture both the Vainshtein recovery
of general relativity and the large distance Yukawa decay \cite{babi10}. For
more details about massive gravity models see also the
following papers \cite{pitt07,babi09,rham11,gong13,babi13}.

So called Yukawa-like fifth force is a framework for deviations from
the inverse-square law in which the gravitational potential deviates
from the usual Newtonian form at large distances due to a
Yukawa-like term in the gravitational potential
\cite{talm88,sere06,card11}. Also it was proposed that the anomalous
observations of the galactic rotation curves could be explained by addition
of a Yukawa correction term to the Newtonian gravity potential \cite{sand84}.
In order to obtain the constraints on Yukawa gravity, studies of the
planetary and stellar orbits at the larger scales, as well as
laboratory searches at the smaller scales have to be performed. Adelberger et al.
\cite{adel09} reviewed experiments with very high sensitivity, placing
constraints about new Yukawa forces from the
exchange of very light scalar, pseudoscalar or vector particles.
Weaker constraints at still smaller scales are available using the
Casimir effect \cite[see e.g.][]{fisc92}. A compilation of
experimental, geophysical and astronomical constraints on Yukawa
violations of the gravitational inverse square law are given in
Figs. 9 and 10 from \cite{adel09} for different ranges. These
results show that the Yukawa term is relatively well constrained for
the short ranges (especially at sub-mm scale), but for long ranges
further tests are needed, and it would be very important to evaluate
parameters of the Yukawa law for whole range of $\Lambda$. A range
around a few thousand AU has not been investigated yet. However, for
longer distances Yukawa corrections have been successfully applied
to clusters of galaxies setting $\delta = +1/3$
\cite{capo07b,capo09b}. The same value of this parameter also gives a
very good agreement between the theoretical and observational
rotation curves of spiral galaxies \cite{card11}.

For now it seems that the only major problem with Yukawa gravity is
in the case of elliptical galaxies where the observations give a
value of $\delta \approx -0.8$ \cite{napo12}, which is inconsistent
with the one previously found for spiral galaxies. Napolitano et al. \cite{napo12}
found a possible explanation for this inconsistency,
according to which $\delta$ might be correlated with the galaxy
anisotropy and the scale parameter, where both elliptical and spiral
galaxies follow the same pattern. In that case, $\delta$ could be
interpreted in terms of physics of the gravitating systems after
their spherical collapse \cite{napo12}.

On the other hand, S-stars are the bright stars which move around
the centre of our Galaxy
\cite{ghez00,scho02,gill09a,gill09b,ghez08,genz10}
where the compact radio source Sgr A$^\ast$ is located. These stars,
together with recently discovered dense gas cloud falling towards
the Galactic Centre \cite{gill12}, provide the most convincing
evidence that Sgr A$^\ast$ represents a massive compact object around which S-stars
are orbiting \cite{genz10}. For one of them, called S2, there are
some observational indications that its orbit could deviate from the Keplerian
case due to relativistic precession \cite{gill09a,meye12}, but the current
astrometric limit is not sufficient to unambiguously confirm such a claim.

The orbital precession can occur due to relativistic effects, resulting
in a prograde pericentre shift or due to a possible extended mass
distribution, producing a retrograde shift \cite{rubi01}. Both
prograde relativistic and retrograde Newtonian pericentre shifts
will result in rosette shaped orbits. Adkins and McDonnell \cite{adki07}
calculated the precession of Keplerian orbits under
the influence of arbitrary central force perturbations. For some
examples, including the Yukawa potential, they presented the results
using hypergeometric functions. Weinberg et al. \cite{wein05}
discussed physical experiments achievable via the monitoring of
stellar dynamics near the massive black hole at the Galactic Centre
with a diffraction-limited, next-generation, extremely large
telescope (ELT). They demonstrated that the lowest order relativistic effects,
such as the prograde precession, could be detectable if the astrometric
precision would reach a few tenths of mas. The astrometric limit for S2 star
orbit today reaches 0.3 mas \cite{frit10}, and some very recent studies provide
more and more evidence that orbit of S2 star is not closing (see e.g. Fig. 2 in \cite{meye12}).

Here we study a possible application of Yukawa gravity within
Galactic Central Parsec, for explaining the observed precession of orbits
of S-stars. We assumed that the motion of these stars could be
described by the gravitational potential around a massive compact central
object (without speculating about its nature, i.e. whether it is a
black hole or not), in any gravity theory which predicts a Yukawa
correction term. For reviews of various theoretical frameworks yielding a
Yukawa-like fifth force, such as braneworld models, scalar-tensor and
scalar-tensor-vector theories of gravity, or studies of topological defects,
see e.g. \cite{krau01,bert05,bert06,moff05,moff06} and references therein.
Other studies of long-range Yukawa-like modifications of gravity conducted with
different astronomical techniques can be found in
\cite{whit01,amen04,reyn05,seal05,shir05,sere06}. Such a phenomenological
approach is in some sense more general than studying the stellar orbits in a
metric of a massive central black hole, since some classes of modified gravity
theories (such as e.g. $f(R)$ theory) predict the black hole metrics which are
equivalent to those obtained in General Relativity.

The present paper is organized as follows: in section 2 we describe our
simulations of stellar orbits in Yukawa gravity potential; the procedure for
fitting the simulated orbits to two independent sets of astrometric
observations of S2 star is described in section 3; our main results
are presented in section 4, and finally, we point out the most important
conclusions of our studies in section 5.

\section{Simulated orbits of S2 star}
\label{sec:Sec2}

As it was already mentioned, Yukawa gravity potential represents a
widely used phenomenological approach to account for possible deviations from
the Newtonian inverse-square law, by introducing the following exponential
(or so called Yukawa-like) modification to the Newtonian gravitational
potential \cite{sand84,capo11b,napo12}:

\begin{equation}
\Phi \left( r \right) = -\dfrac{GM}{(1+\delta)r}\left[ {1 + \delta
e^{- \left(\dfrac{r}{\Lambda} \right)}} \right],
\label{equ01}
\end{equation}

\noindent where $\Lambda$ is an arbitrary parameter (usually
referred to as range of interaction), depending on the typical scale
of the system under consideration and $\delta$ is a universal
constant. For $\delta = 0$ the Yukawa potential reduces to the Newtonian one, as expected.

We simulated orbits of S2 star in the Yukawa gravity potential (\ref{equ01})
and compared the obtained results with two independent sets of observations of
S2 star, obtained by New Technology Telescope/Very Large Telescope (NTT/VLT),
as well as by Keck telescope (see Fig. 1 in \cite{gill09a}), which are publicly
available as the supplementary on-line data to the electronic
version of the paper \cite{gill09a}. The simulated orbits of S2 star were
obtained in the standard way by numerical integration of differential equations
of motion in Yukawa gravitational potential:
\begin{equation}
\mathbf{\dot{r}}=\mathbf{v},\hspace*{0.5cm}
\mu\mathbf{\ddot{r}}=-\triangledown\Phi\left(
\mathbf{r}\right),
\label{2body}
\end{equation}

\noindent where $\mu$ is so called reduced mass in the two-body
problem. In our calculations we assumed that the mass of central
object is $M$ = 4.3 $\times10^6 M_\odot$ and that the
distance to the S2 star is $d_\star$ = 8.3 kpc \cite{gill09a}.
Perturbations from other members of the stellar cluster, as well as
from some possibly existing extended structures composed from visible
or dark matter \cite{zakh07}, were neglected due to simplicity
reasons. The obtained simulated orbits in the Yukawa potential were
compared with the two sets of observations of the S2 star.
Since the integration of the equations (\ref{2body}) results with
coordinates and velocity components in orbital plane (so called true
orbits), the first step is to project them onto the observer's sky
plane (i.e. to calculate the corresponding apparent orbits), in order
to compare them with observed positions. From the theory of binary
stars it is well known that any point $(x, y)$ on the true orbit
could be projected into the point $(x^c, y^c)$ on the apparent orbit
according to (see e.g.\cite{aitk18,smar30}):
\begin{equation}
x^c=l_1 x+l_2 y ,\hspace*{0.5cm} y^c=m_1 x+m_2 y ,
\label{equ10}
\end{equation}
where the expressions for $l_1, l_2, m_1$ and $m_2$ depend on three
orbital elements ($\Omega$ - longitude of the ascending node,
$\omega$ - longitude of pericenter and $i$ - inclination):
\begin{equation}
\begin{array}{l}
l_1=\cos\Omega\cos\omega-\sin\Omega\sin\omega\cos{i}, \\
l_2=-\cos\Omega\sin\omega-\sin\Omega\cos\omega\cos{i}, \\
m_1=\sin\Omega\cos\omega+\cos\Omega\sin\omega\cos{i}, \\
m_2=-\sin\Omega\sin\omega+\cos\Omega\cos\omega\cos{i}. \\
\end{array}
\label{equ11}
\end{equation}
Radial velocity can be calculated from the corresponding true
position $(r,\theta)$ and orbital velocity $(\dot{r},\dot{\theta})$
using the well known expression in polar coordinates
\cite{aitk18}:
\begin{equation}
v_{rad} = \sin i \left[ \sin(\theta + \omega) \cdot \dot{r} + r
\cos (\theta + \omega) \cdot \dot{\theta} \right] .
\label{vrad1}
\end{equation}
\noindent However, in our case it was more convenient to use
the rectangular coordinates $x=r\cos\theta$ and $y=r\sin\theta$ to
calculate the fitted radial velocities:
\begin{equation}
\begin{array}{ll}
v_{rad} =  & \dfrac{\sin i}{\sqrt{x^2 + y^2}} \left[ \sin(\theta +
\omega)
\cdot(x \dot{x} + y \dot{y}) + \right. \\
& \\
 & \left. + \cos (\theta + \omega) \cdot (x \dot{y} - y \dot{x})
\right], \\
\end{array}
\label{vrad2}
\end{equation}
\noindent where $\theta = \arctan \dfrac{y}{x}$.

Since the present paper does not aim to study the Keplerian orbit of
S2 star and since $i$, $\Omega$ and $\omega$ are needed here only for
transforming from true to apparent coordinates, we used the following
values obtained from the same observations \cite{gill09a}:
$i=134^{\circ}.87$, $\Omega=226^{\circ}.53$ and
$\omega=64^{\circ}.98$. One should also take into account that in the
case of orbital precession $\omega$ is in general a function of time,
and therefore it should be treated accordingly during the fitting
procedure. However, based on some theoretical studies of precession
in Yukawa gravity \cite{sand06,iori08,iori08b} one can expect very
slow change of $\omega$ during the observational interval of S2 star,
and since in equations (\ref{equ11}) $\omega$ is used only as an
argument of $\sin$ and $\cos$, fixing it to a constant value could
introduce only negligible errors. Therefore, we assumed it together
with $l_1$, $l_2$, $m_1$ and $m_2$ as constants when projecting true
positions to their corresponding apparent values.

\section{Fitting procedure}
\label{sec:Sec3}

In that way, for each pair of a priori given values of $\delta$ and
$\Lambda$, there are only four unknown parameters which are to be
obtained by fitting: two components of initial position and two
components of initial velocity in orbital plane, corresponding to the
time of the first observation. We varied both $\delta$ and $\Lambda$, and for
each pair of them found the best fit values of S2 star initial conditions
and the corresponding value of $\chi^2$. The fitting itself was performed using
LMDIF1 routine from MINPACK-1 Fortran 77 library which solves the nonlinear
least squares problems by a modification of Marquardt-Levenberg algorithm
\cite{more80}. Unfortunately, $\delta$ and $\Lambda$ could not be
fitted simultaneously with initial position and velocity, most likely
due to inability of the fitting routine to correctly estimate the
corresponding components of Jacobian by a forward-difference
approximation. Therefore, these two parameters were varied in certain
domains, and their best fit estimates were found at the end of the fitting
procedure as those for which the minimum $\chi^2$ was obtained.

We first adopted value of the universal constant $\delta=1/3$
and varied length scale $\Lambda$ in interval from 10 to 10~000 AU
with increment of 10 AU, in order to see whether this value of
$\delta$ could provide the satisfactory fit, as well as to estimate
the most probable value of $\Lambda$ in this case. For each $\Lambda$
we obtained the best fit orbit using the following fitting
procedure:

\begin{enumerate}
\item initial values for true position $(x_0, y_0)$ and orbital
velocity $(\dot{x}_0, \dot{y}_0)$ of S2 star at the epoch
of the first observation are specified;
\item the positions $(x_i, y_i)$ and velocities $(\dot{x}_i,
\dot{y}_i)$ of S2 star along its true orbit are calculated for all
observed epochs by numerical integration of equations of motion
(\ref{2body}) in the Yukawa gravity potential;
\item the corresponding positions $(x_i^c, y_i^c)$ along the apparent
orbit are calculated using the expressions (\ref{equ10}) and
(\ref{equ11}), as well as the corresponding radial velocities
$(v_{rad}^i)$ using (\ref{vrad2});
\item the reduced $\chi^{2}$ of fit is estimated according the
following expression:
\begin{equation}
\chi^{2} =
\dfrac{1}{2N-\nu}{\sum\limits_{i = 1}^N {\left[ {{{\left(
{\dfrac{x_i^o - x_i^c}{\sigma_{xi}}} \right)}^2}
 + {{\left( \dfrac{y_i^o - y_i^c}{\sigma_{yi}}
\right)}^2}} \right]} },
\label{chi2}
\end{equation}
where $(x_i^o, y_i^o)$ is the $i$-th observed position, $(x_i^c,
y_i^c)$ is the corresponding calculated position, $N$ is the number
of observations, $\nu$ is number of unknown parameters (in our case
$\nu=4$), $\sigma_{xi}$ and $\sigma_{yi}$ are uncertainties of
observed positions;
\item the reduced $\chi^{2}$ is minimized and the final values of
initial positions and velocities are obtained.
\end{enumerate}
Finally, we kept the value of $\Lambda$ which resulted with the
smallest value of minimized reduced $\chi^{2}$.

In order to obtain some more general constraints on the
parameters of Yukawa gravity, we also varied both $\delta$ and
$\Lambda$ and studied the simulated orbits of S2 star which give
at least the same or better fits than the Keplerian orbit. For each
pair of these parameters the reduced $\chi^{2}$ of the best fit is
obtained and used for generating the $\chi^{2}$ maps over the
$\Lambda - \delta$ parameter space. These maps are then used to
study the confidence regions in $\Lambda - \delta$ parameter space.

\begin{figure*}
\centering
\includegraphics[width=0.45\textwidth]{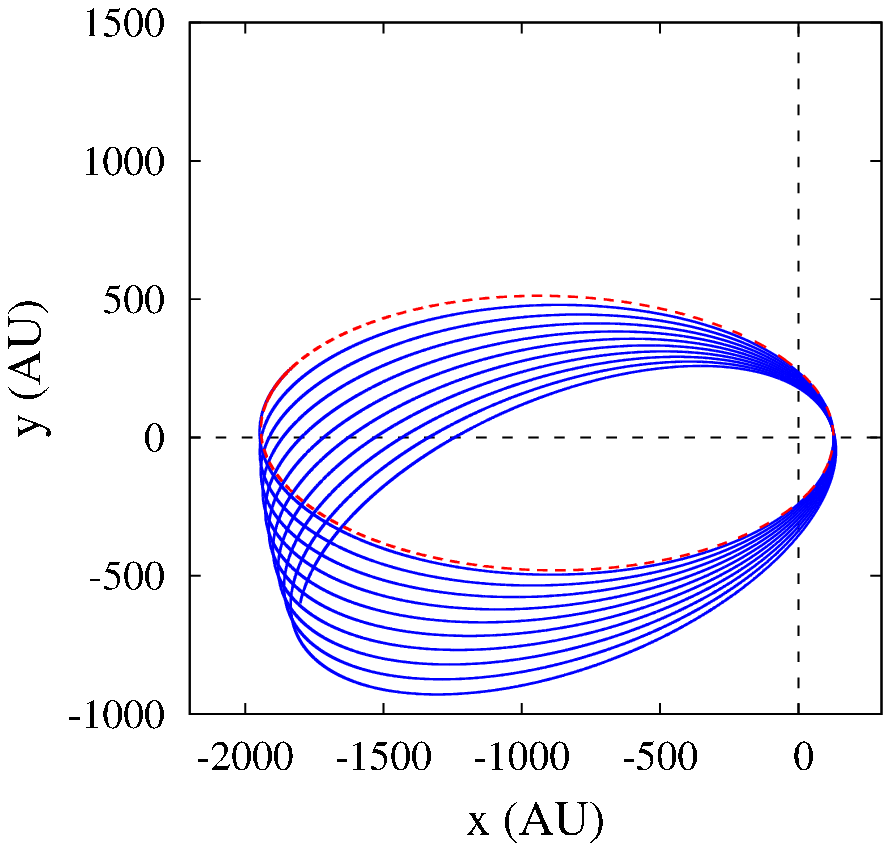}
\hspace*{0.8cm}
\includegraphics[width=0.45\textwidth]{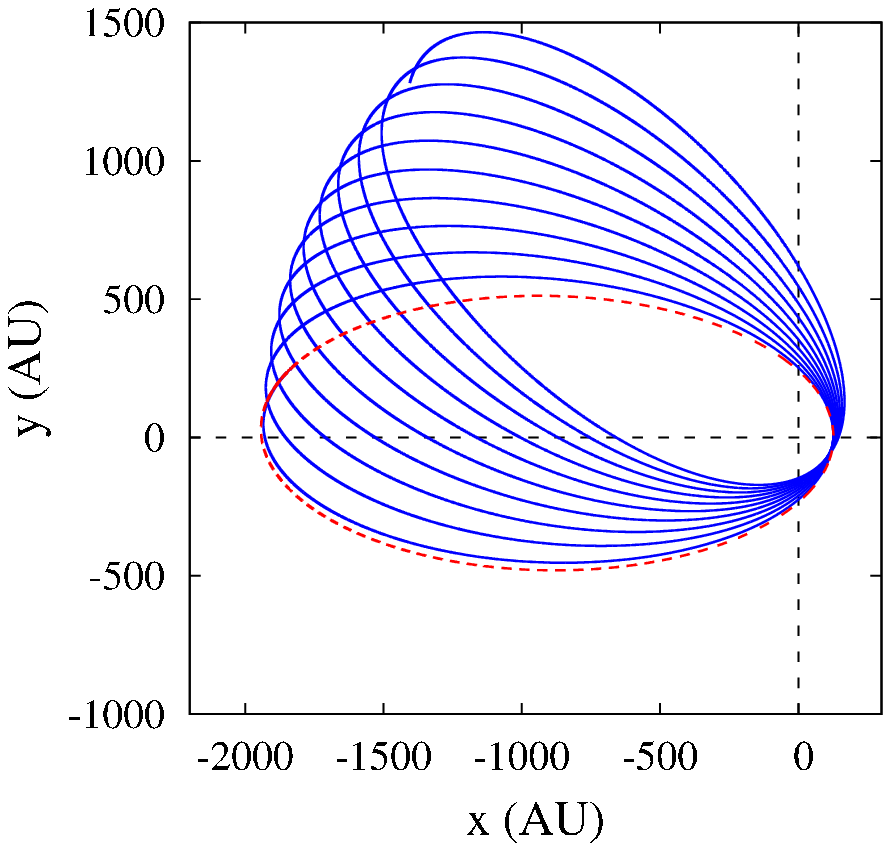}
\caption{Comparisons between the orbit of S2 star in Newtonian
gravity (red dashed line) and Yukawa gravity during 10 orbital
periods (blue solid line) for $\Lambda = 2.59\times 10^3$ AU.
In the left panel $\delta = +1/3$, and in the right $\delta =
-1/3$.}
\label{fig01}
\end{figure*}

\begin{figure*}
\centering
\includegraphics[width=0.45\textwidth]{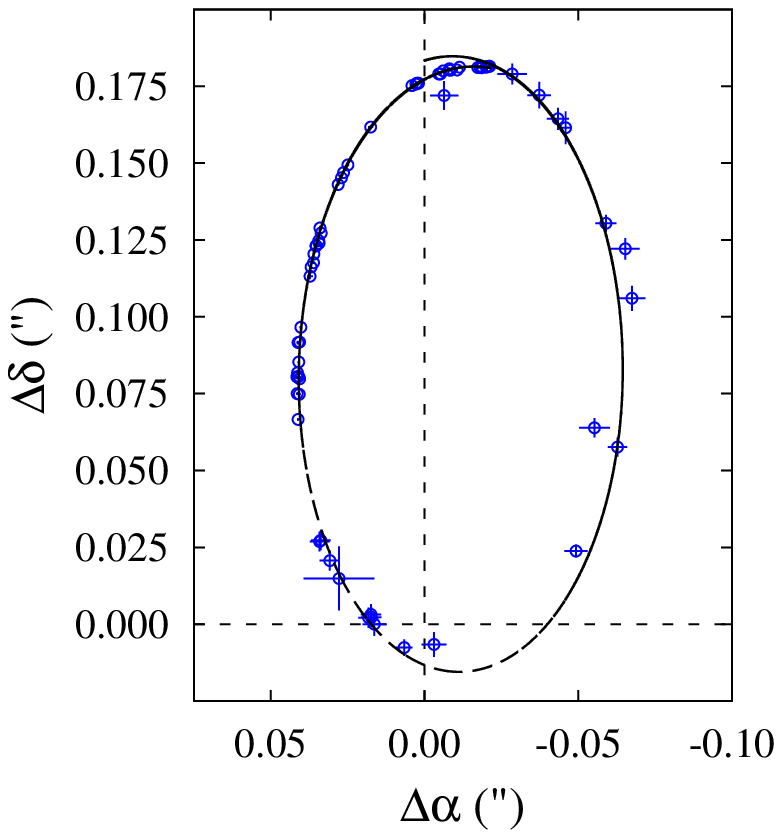}
\hspace*{1cm}
\includegraphics[width=0.45\textwidth]{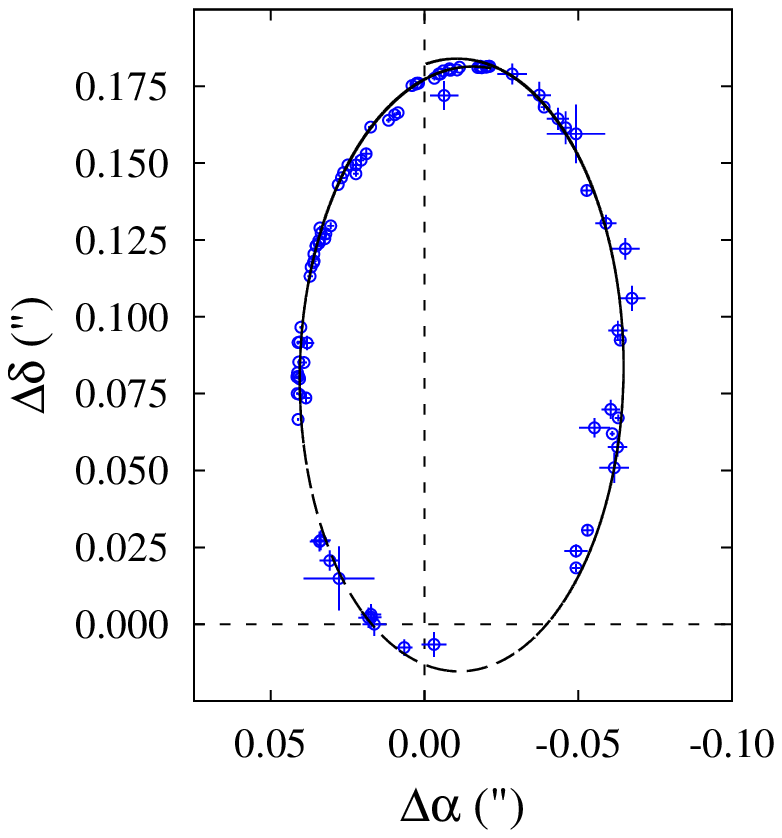}
\caption{The fitted orbits in Yukawa gravity for $\delta = +1/3$
through the astrometric observations of S2 star (denoted by
circles), obtained by NTT/VLT alone (left panel) and NTT/VLT+Keck
(right panel). The best fits are obtained for $\Lambda = 2.59\times
10^3$ AU and $\Lambda = 3.03\times 10^3$ AU, respectively.}
\label{fig02}
\end{figure*}

\begin{figure*}
\centering
\includegraphics[width=0.45\textwidth]{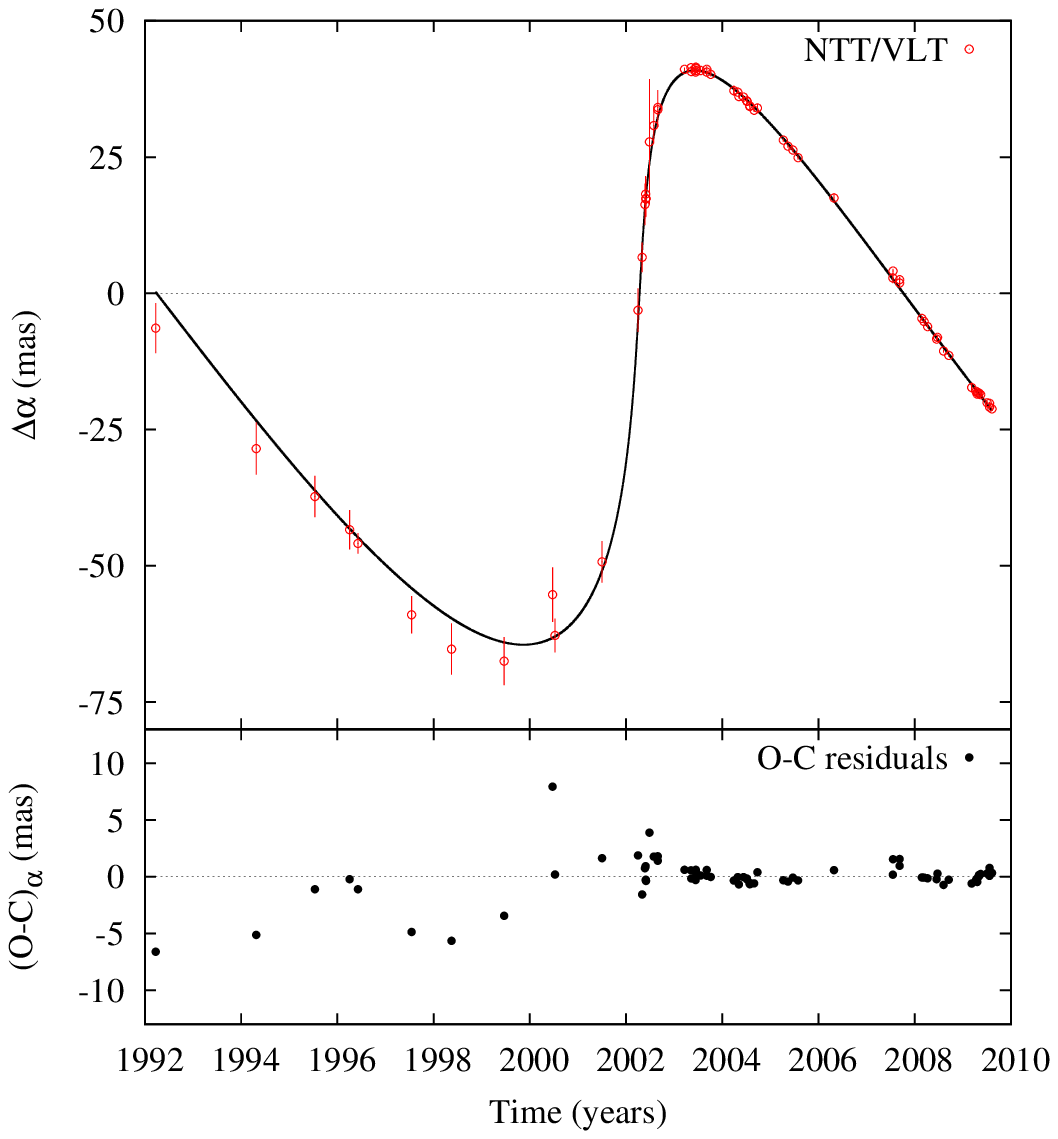}
\hspace*{1cm}
\includegraphics[width=0.45\textwidth]{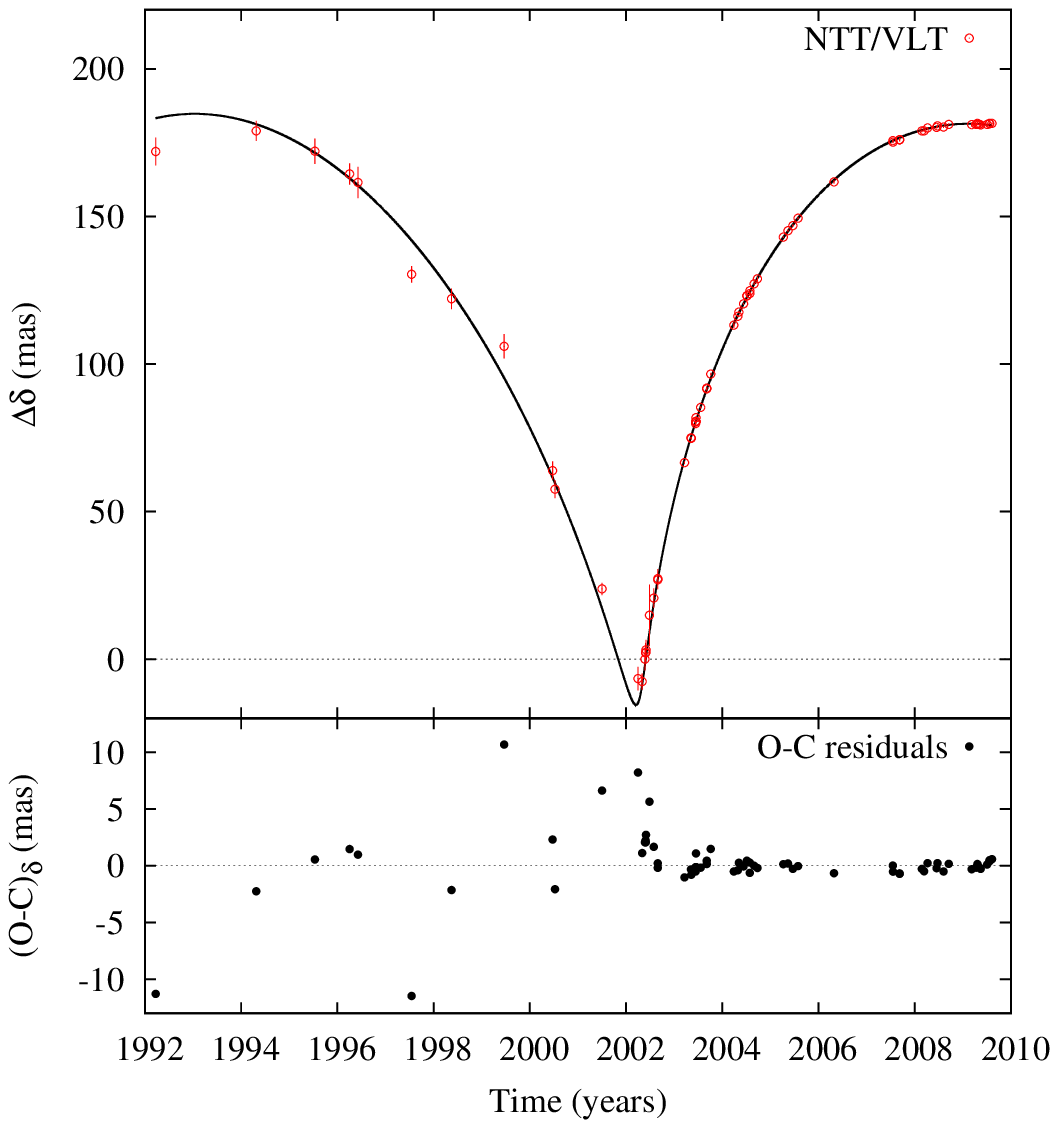}
\caption{The comparisons between the observed (open circles with
error bars) and fitted (solid lines) coordinates of S2 star (top), as
well as the corresponding O-C residuals (bottom). The left panel
shows the results for $\Delta\alpha$ and right panel for
$\Delta\delta$ in the case of NTT/VLT observations and Yukawa
gravity potential with $\delta = +1/3$ and $\Lambda = 2.59\times
10^3$ AU.}
\label{fig03}
\end{figure*}

\begin{figure*}
\centering
\includegraphics[width=0.45\textwidth]{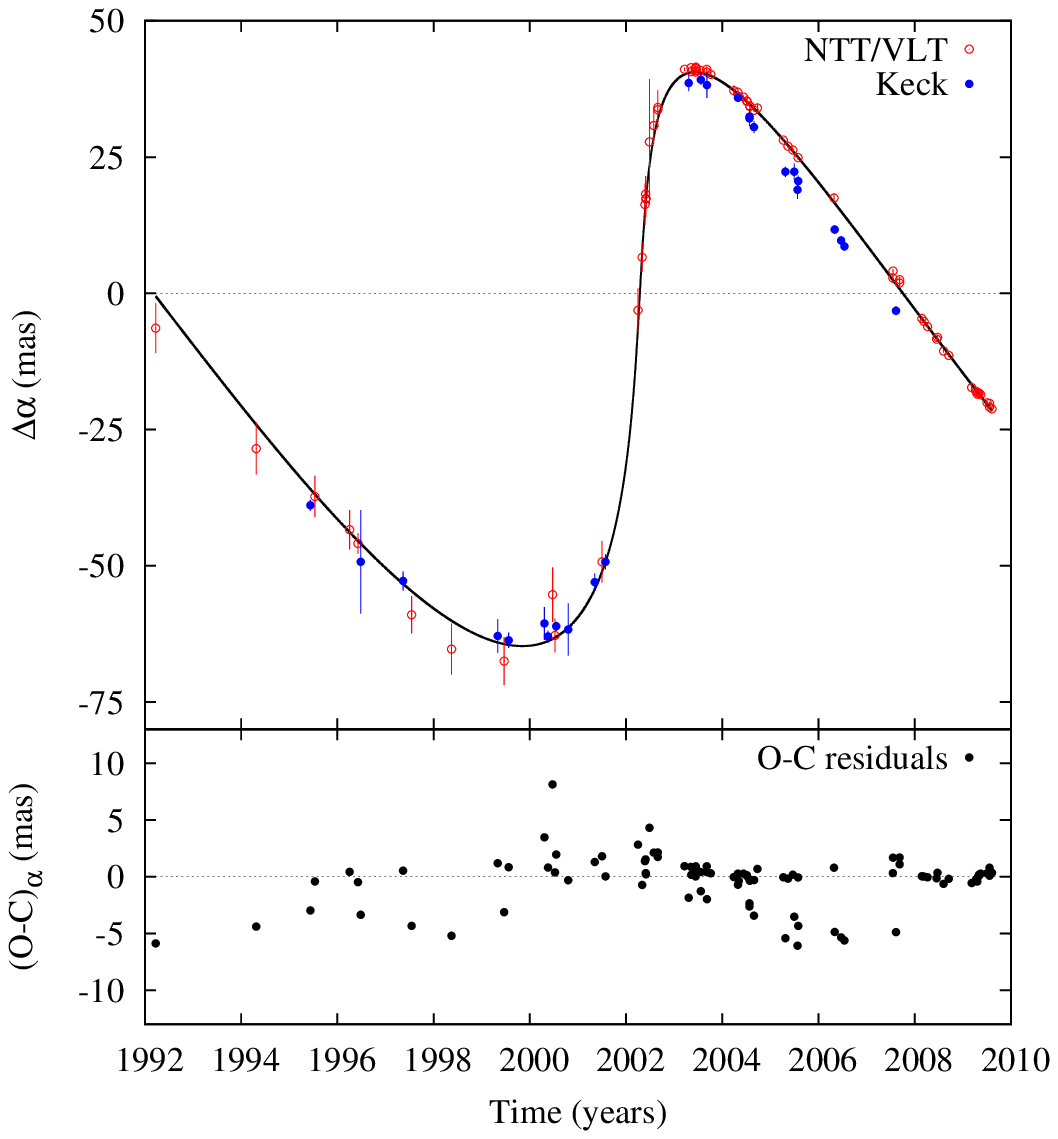}
\hspace*{1cm}
\includegraphics[width=0.45\textwidth]{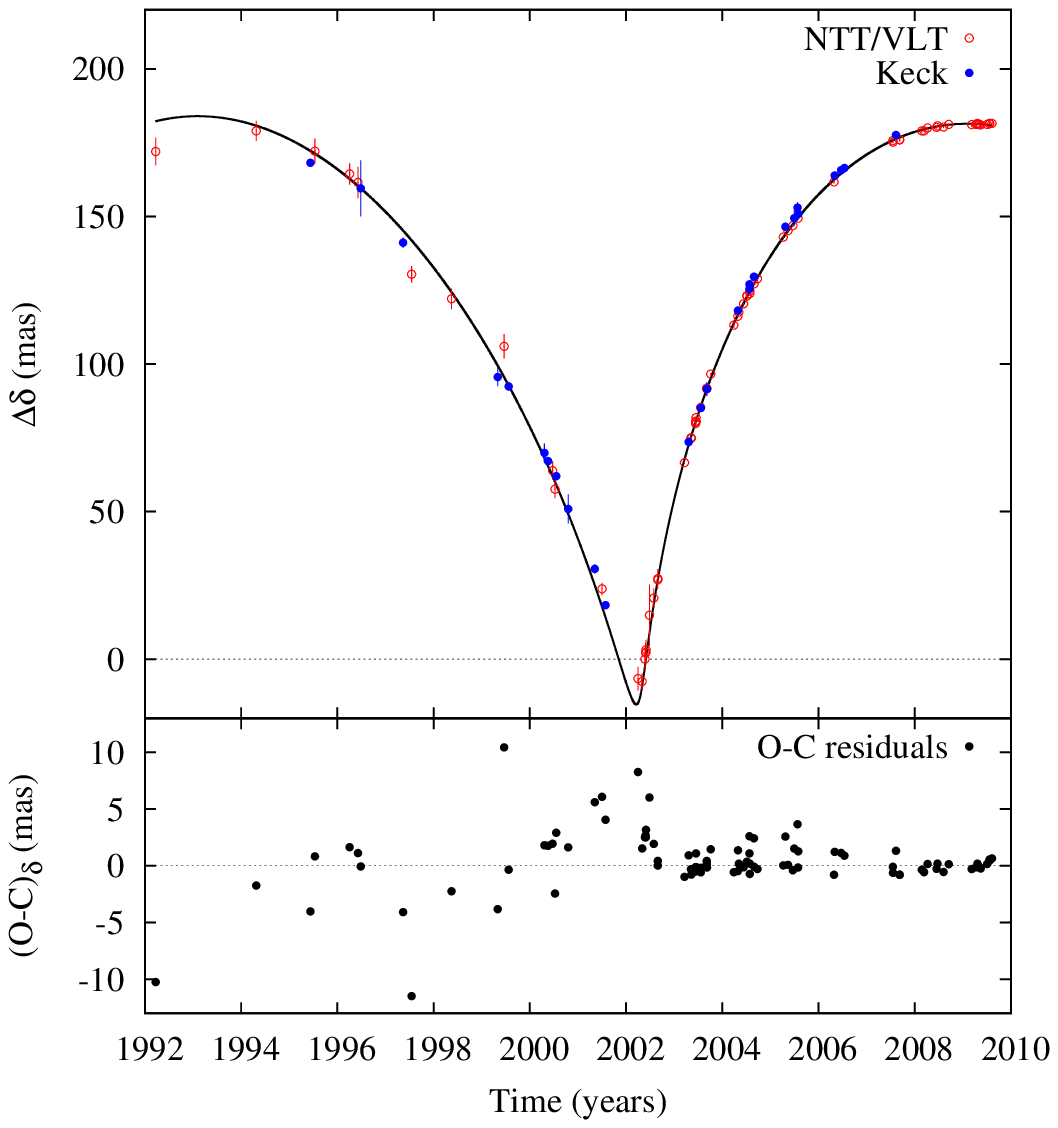}
\caption{The same as in Fig. \ref{fig03}, but for NTT/VLT+Keck
combined observations and for Yukawa gravity potential with $\Lambda
= 3.03\times 10^3$ AU.}
\label{fig04}
\end{figure*}

\begin{figure*}
\centering
\includegraphics[width=0.45\textwidth]{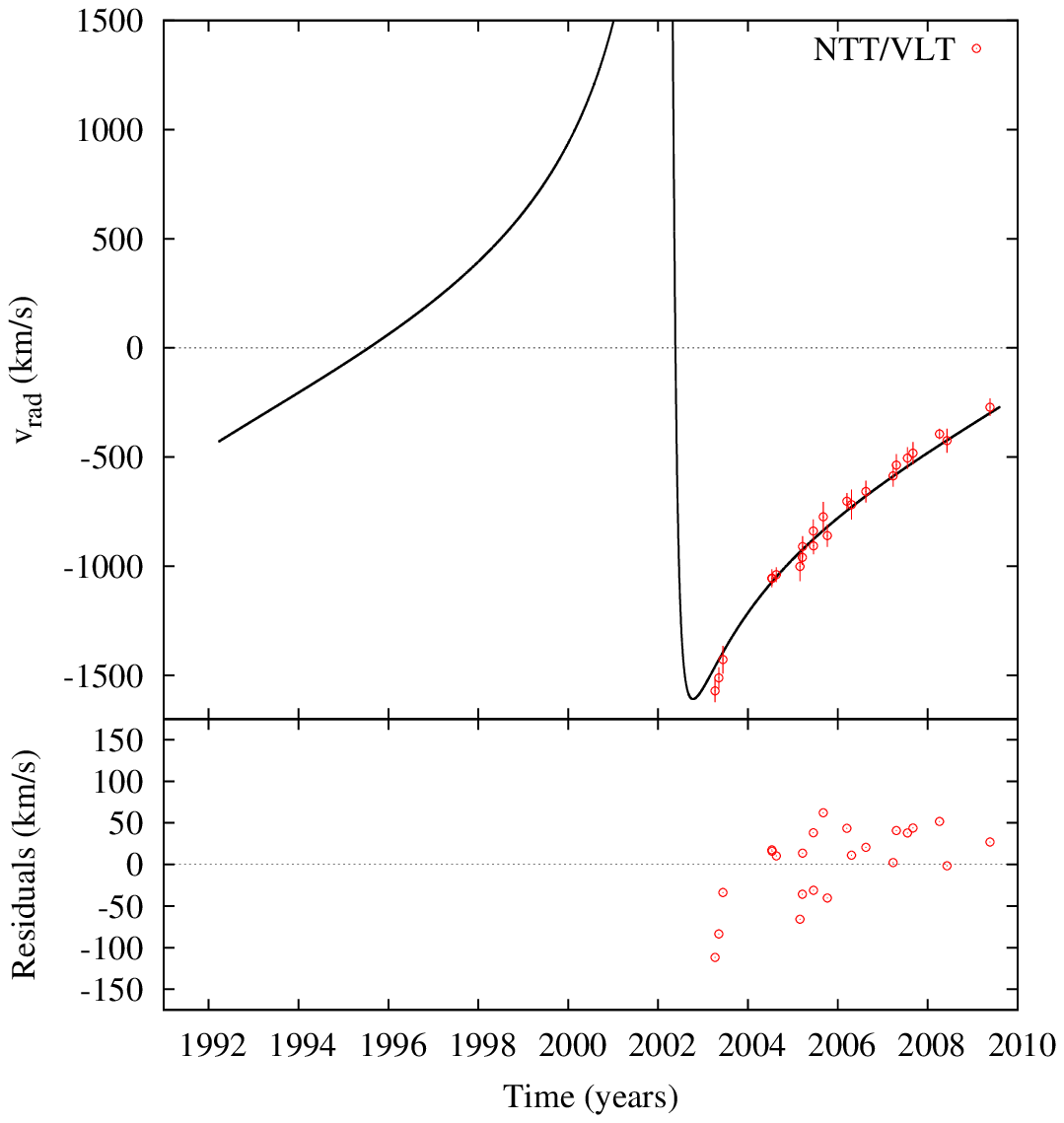}
\hspace*{1cm}
\includegraphics[width=0.45\textwidth]{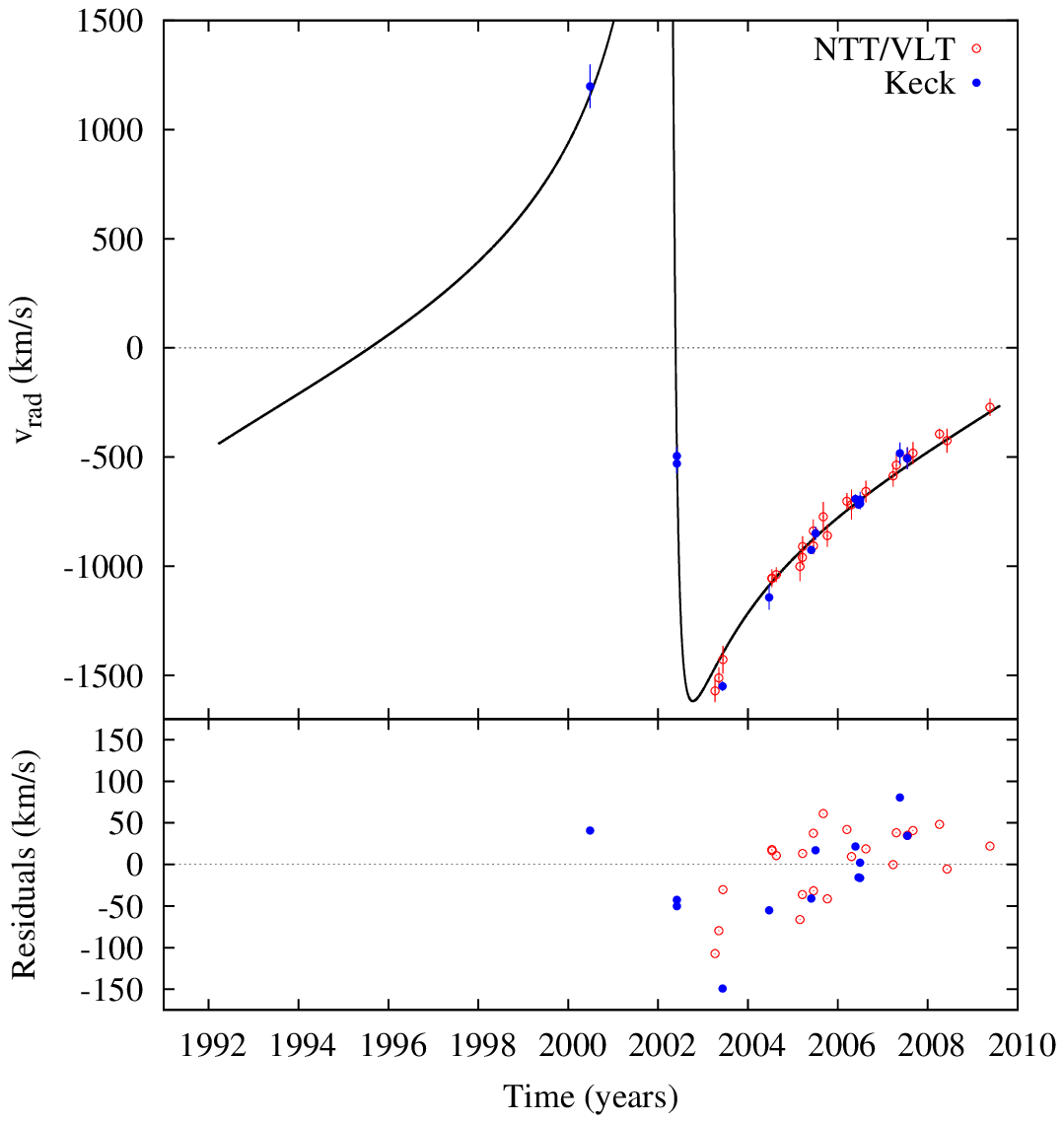}
\caption{The comparisons between the observed (circles with error
bars) and fitted (solid lines) radial velocities of S2 star (top), as
well as the corresponding O-C residuals (bottom). The left panel
shows the results in the case of NTT/VLT observations and Yukawa
gravity potential with $\Lambda = 2.59\times 10^3$ AU, while the
right panel shows the results for NTT/VLT+Keck combined observations
and for Yukawa gravity potential with $\Lambda = 3.03\times 10^3$ AU.
In both cases $\delta = +1/3$.}
\label{fig05}
\end{figure*}

\begin{figure*}
\centering
\includegraphics[width=0.45\textwidth]{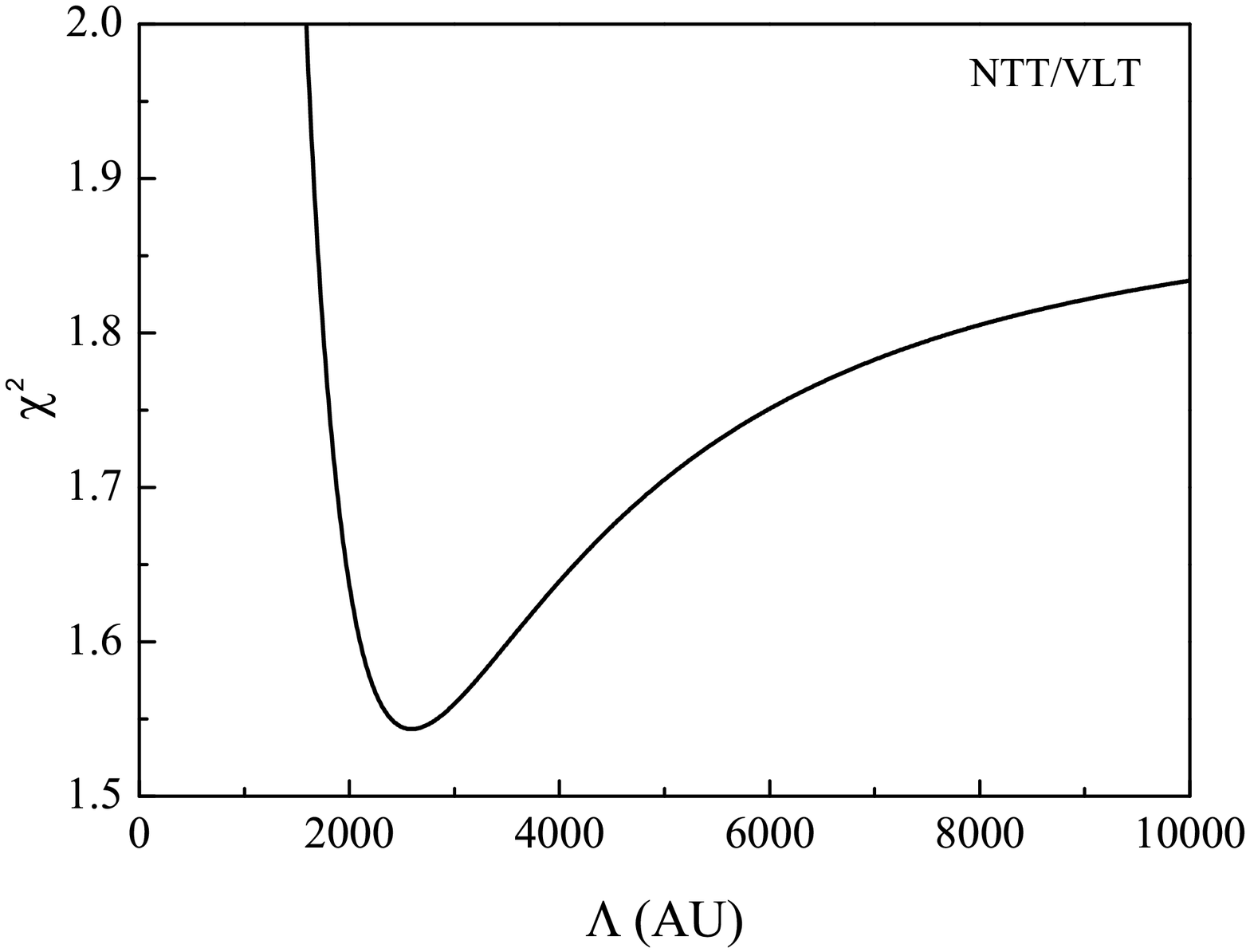}
\hspace*{1cm}
\includegraphics[width=0.45\textwidth]{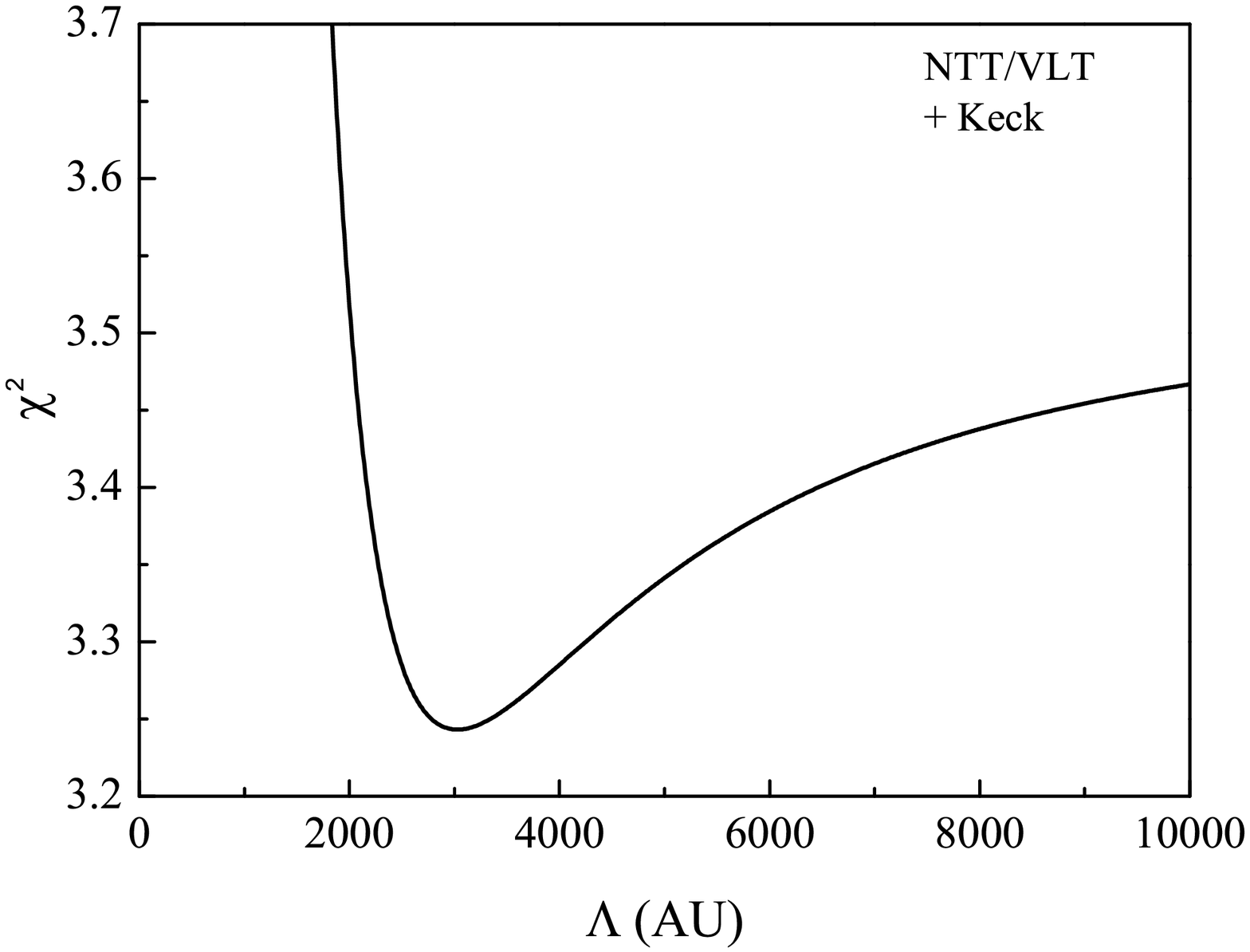}
\caption{The reduced $\chi^{2}$ for $\delta$=1/3 as a function of
$\Lambda$ in case of  NTT/VLT alone (left) and combined NTT/VLT+Keck
(right) observations.}
\label{fig06}
\end{figure*}

\begin{figure*}
\centering
\includegraphics[width=0.45\textwidth]{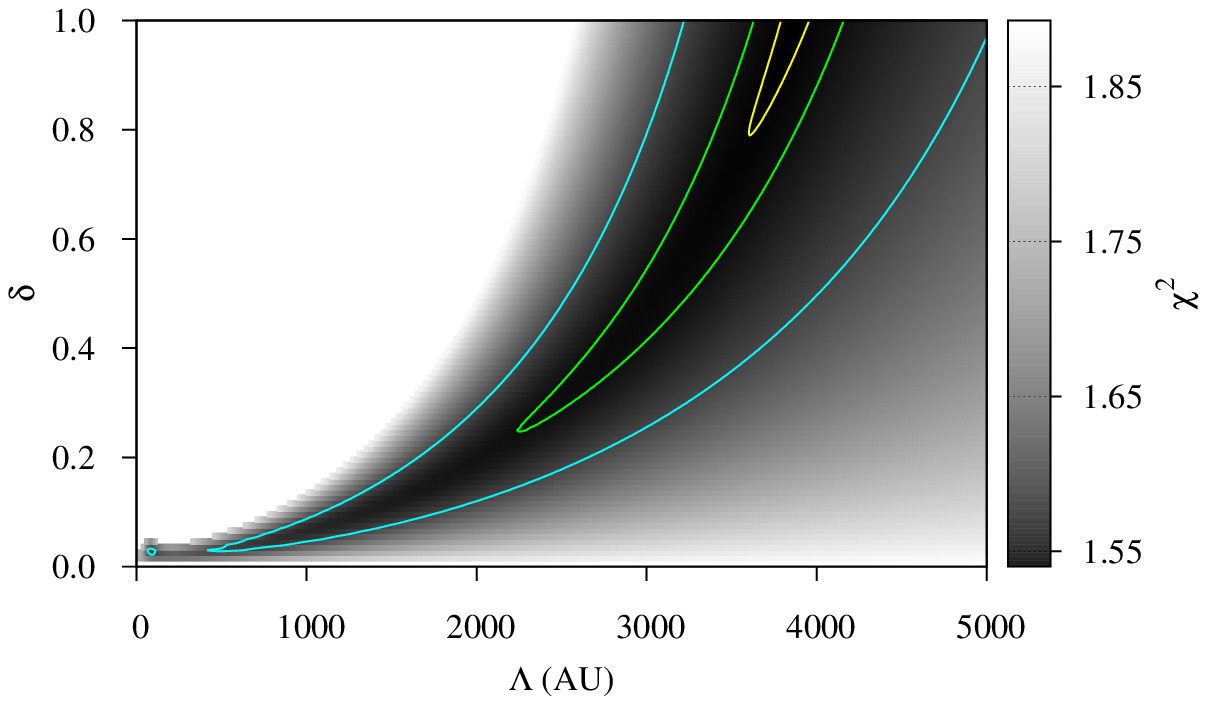}
\hspace*{1cm}
\includegraphics[width=0.45\textwidth]{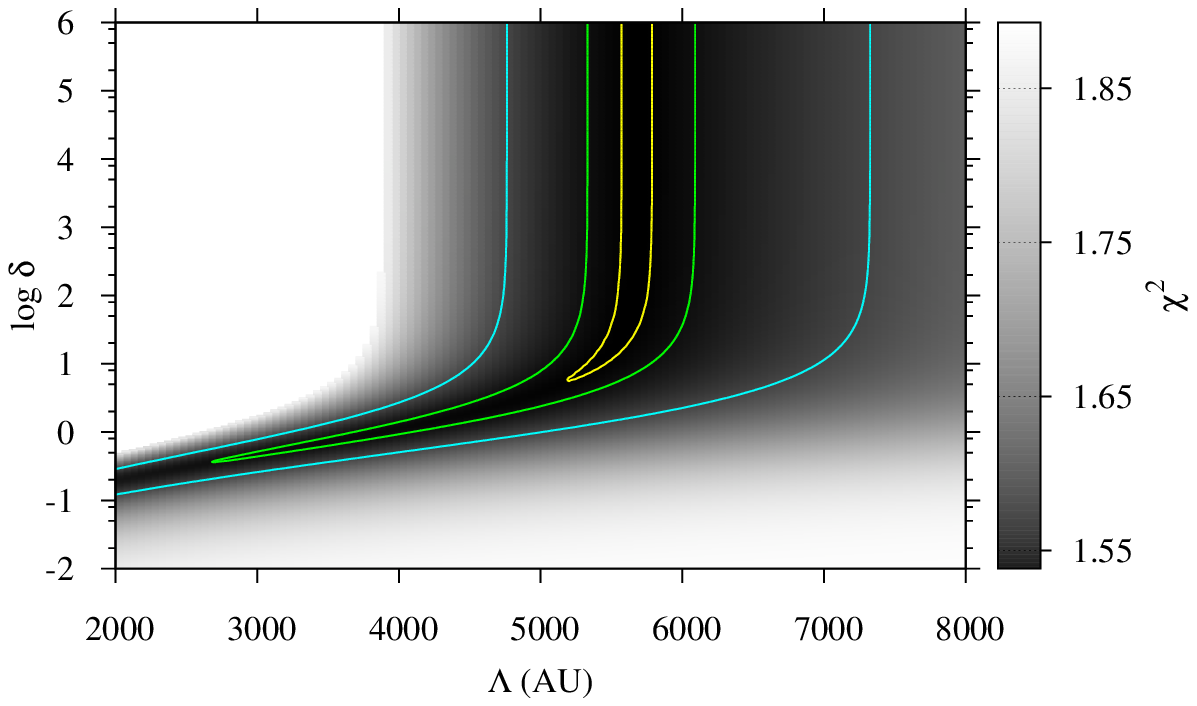}
\caption{The maps of reduced $\chi^2$ over the $\Lambda -
\delta$ parameter space in case of NTT/VLT observations. The left
panel corresponds to $\delta \in [0, 1]$, and the right panel to the
extended range of $\delta \in [0.01, 10^6]$. The shades of gray
color represent the values of the reduced $\chi^2$ which are less
than the corresponding value in the case of Keplerian orbit, and
three contours (from inner to outer) enclose the confidence regions
in which the difference between the current and minimum reduced
$\chi^2$ is less than 0.0005, 0.005 and 0.05, respectively.}
\label{fig07}
\end{figure*}

\begin{figure*}
\centering
\includegraphics[width=0.45\textwidth]{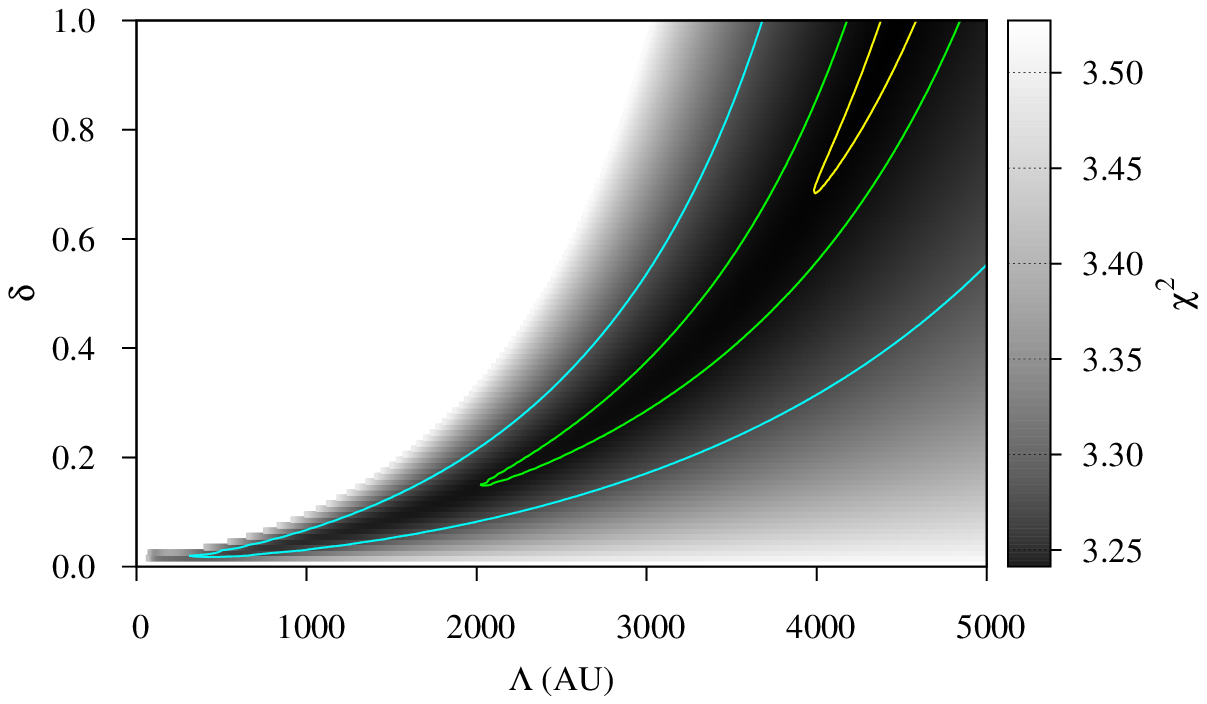}
\hspace*{1cm}
\includegraphics[width=0.45\textwidth]{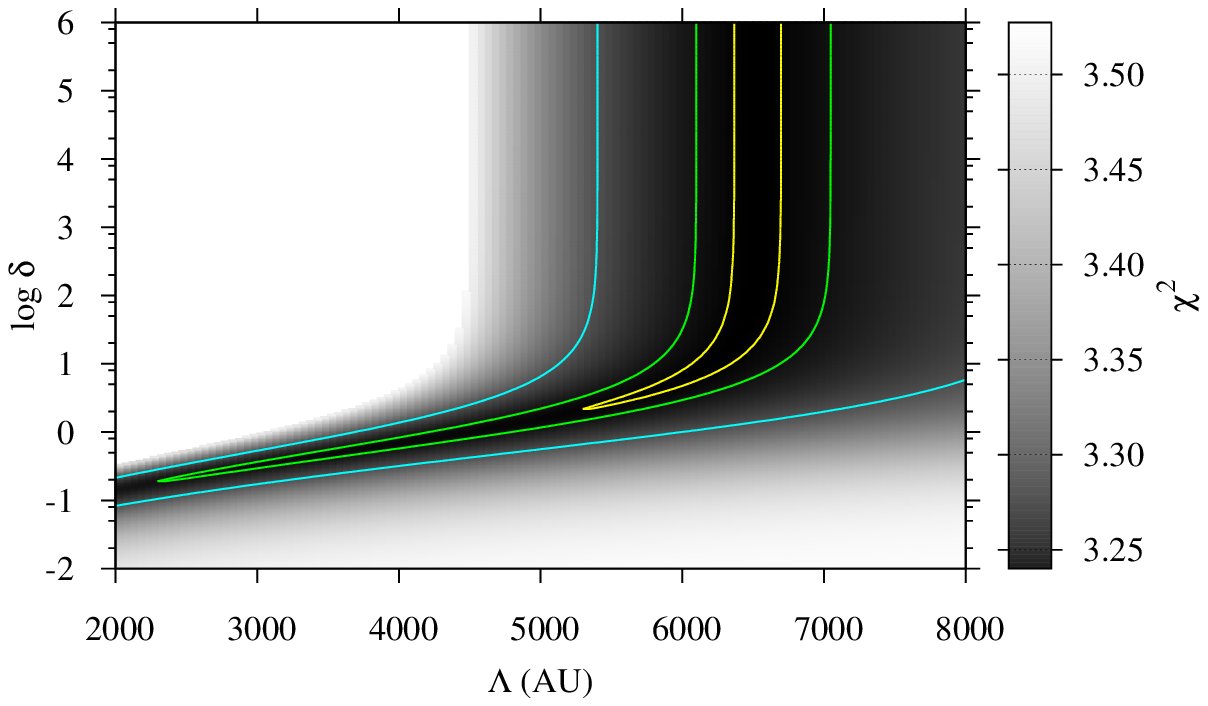}
\caption{The same as in Fig. \ref{fig07}, but for the
combined NTT/VLT+Keck observations.}
\label{fig08}
\end{figure*}

\section{Results: simulations vs observations}
\label{sec:Sec4}

The simulated orbits of S2 star around the central object in Yukawa
gravity (blue solid line) and in Newtonian gravity (red dashed line)
for $\Lambda = 2.59\times 10^3$ AU and $\delta = +1/3$ (left panel)
and $\delta = -1/3$ (right panel) during 10 periods, are presented
in Fig. \ref{fig01}. We can notice that for $\delta = -1/3$
the precession has the negative direction and when $\delta = +1/3$ the
precession has the positive direction. Our analysis shows that the Yukawa
gravity potential induces precession of S2 star orbit in the same
direction as General Relativity for $\delta > 0$ and for $\delta <
-1$, and in the opposite direction for $-1 <\delta < 0$ as in the
case of extended mass distribution or in $R^n$ gravity
\cite{bork12}.

We used these simulated orbits to fit the observed orbits of S2 star.
The best fit (according to NTT/VLT data) is obtained for the scale
parameter: $\Lambda = 2.59\times 10^3$ AU, for which even a
significant strength of the Yukawa interaction could be expected
according to the planetary and Lunar Laser Ranging constraints
\cite{adel09}. Lunar (and artificial satellite) Laser ranging (LLR)
is one of the most accurate techniques to test gravitational physics and
Einstein's theory of General Relativity \cite{ciuf10,xie10,ciuf11} at the
corresponding length scales. In particular, LLR has provided very accurate
tests of the strong equivalence principle at the foundations of General
Relativity and of the weak equivalence principle, at the basis of any metric
theory of gravity. Also, it has provided strong limits to the values of the
so-called PPN (Parametrized Post-Newtonian) parameters \cite{ciuf11}.
The Solar System and LLR constraints on the range of Yukawa interaction are
shown in Fig. 16 (see also Table 8) in Adelberger et al. \cite{adel09},
according to which $\Lambda \gg 1.5\times 10^{11}$ m and $\Lambda \gg 4\times 10^8$ m,
respectively, and thus is in accordance with our findings.

In Fig. \ref{fig02} we presented two comparisons between the fitted
orbits in Yukawa gravity for $\delta = +1/3$ through the astrometric
observations of S2 star by NTT/VLT alone (left) and NTT/VLT+Keck
combination (right). In order to combine NTT/VLT and Keck data sets,
the position of the origin of Keck observations is first shifted by
$\Delta x=3.7$ and $\Delta y=4.1$ mas, following the suggestion given
in \cite{gill09a}. In the first case the best fit is obtained for
$\Lambda = 2.59\times 10^3$ AU, resulting with reduced $\chi^2=1.54$,
and in the second case for $\Lambda = 3.03\times 10^3$ AU with
reduced $\chi^2=3.24$. As one can see from these figures, in both
cases there is a good agreement between the theoretical orbits and
observations, although the higher value of reduced $\chi^2$ in the
second case indicates possibly larger positional difference between
the two coordinate systems, as also noted in \cite{gill09a}.
These figures also show that the simulated orbits of S2 are not
closed in vicinity of apocenter, indicating a possible orbital
precession.

In Figs. \ref{fig03} and \ref{fig04} we presented the comparisons
between the observed and fitted coordinates of S2 star and their O-C
residuals in the case of NTT/VLT observations, as well as
NTT/VLT+Keck combined data set, respectively. One can notice that in
both cases, O-C residuals are higher in the first part of observing
interval (up to the 12 mas) and much less in its second part (less
than 2 mas). Due to adopted merit function given by Eq.(\ref{chi2}),
our fitting procedure assigned greater weight to these
latter, more precise observations. Also, the O-C residuals are larger
in the case of the combined NTT/VLT+Keck observations most likely due
to the shift of the origin of the coordinate system, which was necessary in
order to get a reasonable fit. That is why we also presented the results for NTT/VLT
measurements alone.

We also made the comparisons between the observed and fitted
radial velocities of S2 star (see Fig. \ref{fig05}) for NTT/VLT
data alone (left) and  NTT/VLT+Keck combination (right). In the
bottom parts of both panels in Fig. \ref{fig05} the best fit O-C
residuals for radial velocities are also given. As it can be seen
from Fig. \ref{fig05}, we also obtained satisfactory agreement
between the predicted and observed radial velocities of S2 star.

Figure \ref{fig06} presents the reduced $\chi^{2}$ for all fits with
fixed value of $\delta$=1/3 as a function of the other parameter of
Yukawa gravity $\Lambda$ which was varied from 10 to 10 000 AU.
In the case of NTT/VLT observations the minimum of reduced $\chi^{2}$
is 1.54 and is obtained for $\Lambda = 2.59\times 10^3$ AU, while in
the case of NTT/VLT+Keck combined data set the minimal value of 3.24
is obtained for $\Lambda = 3.03\times 10^3$ AU. For both cases the
reduced $\chi^{2}$ for Keplerian orbits ($\delta=0$) are 1.89 and
3.53, respectively, and thus significantly higher than the
corresponding minima for $\delta=1/3$. This means that Yukawa gravity
describes observed data even better than Newtonian gravity and that
$\delta=1/3$ is valid value at galactic scales.

Figs. \ref{fig07} and \ref{fig08} present the maps of the reduced
$\chi^{2}$ over the $\Lambda - \delta$ parameter space for all
simulated orbits of S2 star which give at least the same or better
fits than the Keplerian orbits. These maps are obtained by the same
fitting procedure as before. The left panels of both figures
correspond to $\delta \in [0, 1]$ and $\Lambda[\mathrm{AU}] \in [10,
5000]$, and the right panels to the extended range of $\delta \in
[0.01, 10^6]$ and $\Lambda[\mathrm{AU}] \in [2000, 8000]$.
Three contours (from inner to outer) enclose the confidence
regions in which the difference between the current and minimum
reduced $\chi^2$ is less than 0.0005, 0.005 and 0.05, respectively.
As it can be seen from Fig. \ref{fig07}, the most probable
value for the scale parameter $\Lambda$, in the case of NTT/VLT
observations of S2 star, is around 5000 - 6000 AU, while in the case
of NTT/VLT+Keck combined data set (Fig. \ref{fig08}), the most
probable value for $\Lambda$ is around 6000 - 7000 AU. In both cases
$\chi^2$ asymptotically decreases as a function of $\delta$, and
hence, it is not possible to obtain reliable constrains on the
universal constant $\delta$ of Yukawa gravity. Also, these
two parameters $\delta$ and $\Lambda$ are highly correlated in the
range $(0 <\delta < 1)$. For $\delta > 2$ (the vertical strips) they
are not correlated.

As it could be also seen from left panels of Figs. \ref{fig07} and
\ref{fig08}, the values $\delta\approx 1/3$ result with very good
fits for which the reduced $\chi^2$ deviate from the minimal value
for less than 0.005 (middle contours in both figures). The
corresponding values for $\Lambda$ range approximately from 2500 to
3000 AU.  For $\delta=1/3$ we obtained the following values: $\Lambda
= 2590 \pm 5$ AU (NTT/VLT data) and $\Lambda = 3030 \pm 5$ AU
(NTT/VLT+Keck combined data).

Although both observational sets indicate that the orbit of S2
star most likely is not a Keplerian one, the current astrometric
limit is not sufficient to unambiguously confirm such a claim.
However, the accuracy is constantly improving from around 10 mas
during the first part of the observational period, currently
reaching around 0.3 mas. We hope that in the future, it will be
possible to measure the stellar positions with much better accuracy
of $\sim 10\ \mu$as \cite{gill10}.

\section{Conclusions}
\label{sec:Sec5}

In this paper orbit of S2 star has been investigated in the
framework of the Yukawa gravity. Using the observed positions of S2
star around the Galactic Centre we constrained the parameters of Yukawa
gravity. Our results show that:
\begin{enumerate}
\item the most probable value for Yukawa gravity parameter $\Lambda$ in the
case of S2 star, is around 5000 - 7000 AU and that the current observations do
not enable us to obtain the reliable constraints on the universal constant
$\delta$;
\item the same universal constant $\delta$ which was successfully
applied to clusters of galaxies
\citep{capo07b,capo09b} and rotation curves of spiral galaxies
\citep{card11} also gives a good agreement in the case of observations
of S2 star orbit;
\item the scale parameter of Yukawa gravity in the case of S2 star
for $\delta = +1/3$ is about: $\Lambda \approx 3000 \pm 1500$ AU;
\item for vanishing $\delta$, we recover the Keplerian orbit
of S2 star;
\item for $\delta = +1/3$ there is orbital precession in positive
direction like in General Relativity, and for $\delta = -1/3$ the
precession has negative direction, as in the case of extended mass
distribution or in $R^n$ gravity \cite{bork12};
\item the two parameters of Yukawa gravity are highly correlated in the range
$(0 <\delta < 1)$. For $\delta > 2$ they are not correlated.
\end{enumerate}

Borka et al. \cite{bork12} found that $R^n$ gravity may not represent
a good candidate to solve both the rotation curves problem of spiral
galaxies and the orbital precession of S2 star for the same value of
the universal constant $\beta$ ($\beta$=0.817 and $\beta\sim$0.01,
respectively). According to the above results, the opposite
conclusion is not eliminated in the case of Yukawa gravity with
$\delta=1/3$.

The constraints on parameter $\Lambda$ obtained in the present paper
are in agreement with the corresponding Solar System and LLR constraints
presented by Adelberger et al. \cite{adel09}, according to which $\Lambda \gg
1.5\times 10^{11}$ m and $\Lambda \gg 4\times 10^8$ m, respectively. Also, the
constraints on the Yukawa-like modifications of gravitation from Solar System
planetary motions showed that one can assume values of the parameter $\Lambda$
on the same order of magnitude, or larger than the typical sizes of the
planetary orbits in the Solar System \cite{iori07}. This is common assumption in
many modified theories of gravity \cite{iori07, moff07} and is also in
accordance with our findings.

However, one should keep in mind that we considered an
idealized model ignoring many uncertain factors, such as an extended
mass distribution, perturbations from nonsymmetric mass distribution,
etc. Therefore, the future observations with advanced facilities,
such as GRAVITY which will enable extremely accurate measurements of
the positions of stars of $\sim$ 10 $\mu$as \cite{gill10}, or E-ELT
with expected accuracy of $\sim$ 50-100 $\mu$as \cite{eelt}, are
needed in order to verify these claims.

\acknowledgments

This research is part of the project 176003 ''Gravitation and the
large scale structure of the Universe'' supported by the Ministry of
Education, Science and Technological Development of the Republic of
Serbia. AFZ was supported in part by the NSF (HRD-0833184) and NASA
(NNX09AV07A) at NCCU in Durham. The authors thank a referee for a
constructive criticism.


\end{document}